\documentclass[aps,prb,twocolumn,superscriptaddress,10pt]{revtex4-1}

\usepackage{graphicx}
\usepackage{amsmath}
\usepackage{amssymb}
\usepackage{amsfonts}
\usepackage{bm}
\usepackage{psfrag}
\usepackage{pstricks}
\usepackage{color}

\pdfoutput=1

\begin{document}
\title{Fully analytic valence force field model for the elastic and inner elastic properties of diamond and zincblende crystals}

\author{Daniel S.~P.~Tanner}
\email{danielsptanner@gmail.com}
\affiliation{Tyndall National Institute, Lee Maltings, Dyke Parade, Cork T12 R5CP, Ireland}

\author{Miguel~A. Caro}
\affiliation{Department of Electrical Engineering and Automation, Aalto University, Espoo 0215, Finland} 
\affiliation{Department of Applied Physics, Aalto University, Espoo 0215, Finland}

\author{Stefan Schulz}
\affiliation{Tyndall National Institute, Lee Maltings, Dyke Parade, Cork T12 R5CP, Ireland}

\author{Eoin P.~O'Reilly}
\affiliation{Tyndall National Institute, Lee Maltings, Dyke Parade, Cork T12 R5CP, Ireland}
\affiliation{Department of Physics, University College Cork, Cork T12 YN60, Ireland}

\begin{abstract}
Using a valence force field model based on that introduced by Martin, we present
three related methods through which we analytically determine valence
force field parameters. The methods introduced allow easy derivation of valence
force field parameters in terms of the Kleinman parameter $\zeta$ and bulk properties
of zincblende and diamond crystals. We start with a model suited for covalent and weakly ionic
materials, where the valence force field parameters are derived in terms of $\zeta$
and the bulk elastic constants $C_{11}$, $C_{12}$, and $C_{44}$. We show that this
model breaks down as the material becomes more ionic and specifically when the 
elastic anisotropy factor $A = 2C_{44}/(C_{11}-C_{12}) > 2$. The analytic model
can be stabilised for ionic materials by including Martin's electrostatic 
terms with effective cation and anion charges in the valence force field model.
Inclusion of effective charges determined via the optical phonon mode splitting
provides a stable model for all but two of the materials considered (zincblende GaN and AlN).
A stable model is obtained for all materials considered by also utilising the inner elastic constant
$E_{11}$ to determine the magnitude of the effective charges used in the Coulomb interaction. 
Test calculations show that the models describe well structural relaxation in superlattices 
and alloys, and reproduce key phonon band structure features. 

\end{abstract}

\date{\today}


\pacs{78.67.De, 73.22.Dj, 73.21.Fg, 77.65.Ly, 73.20.Fz, 71.35.-y}

\maketitle

\section{Introduction}
The use of interatomic potentials for the study of the elastic properties of solids has a long history.
Relations between the elastic constants of crystals were obtained as early as the 19th century,\cite{Saint-Ven} when
the Cauchy relations were derived analytically using a simple central pairwise atomic interaction.
Later it was found by Born that in order to model covalent crystals, whose elastic constants do
not bear such simple relation to each other, non-central interatomic interactions needed to be included.\cite{Born_dynamic} 
In this manner the complexity of the involved potentials grew as the variety of systems to 
which they were applied grew, with today's 
interatomic potentials involving up to hundreds of different atomic interactions, 
parameterised and implemented at great computational expense.\cite{ThSw15,BaAl10}
 
Generally, interatomic potentials were used to predict unknown crystal properties from known ones. 
For example, in the early literature, interatomic potentials parameterised from known elastic 
constants (for example $C_{11}$ and $C_{12}$) or phonon frequencies were used to predict such 
experimentally inaccessible quantities as inner elastic constants and internal 
strain,\cite{Keating66,Born_dynamic, CousinsInner,Cousins82} temperature, pressure and 
strain dependence of elastic constants,\cite{FeKl69,ElFa02,CousinsPhysBI} third order 
elastic constants,\cite{Keating66,CousinsPhysBI} vibrational properties,\cite{MusPop62,McSo67,NuBi67} 
as well as general insights into interatomic forces, and explanations for trends 
in elastic properties.\cite{Keating66,Ma1970, BazantThesis}  
 
More recently, with the advent of ab-initio calculations, capable of determining all \textit{bulk} elastic, inner elastic
and dynamical properties of a crystal to a high accuracy, the use of interatomic potentials for
the prediction of the properties of simple bulk systems has dropped off:
while the predictions of interatomic potentials were useful first approximations, their use was no longer
justified when such properties could be easily calculated to a high accuracy using first-principles methods. 
Furthermore, properties such as the previously experimentally inaccessible internal relaxation 
and various shear moduli, which were formerly predicted by interatomic potentials, 
may now be used in their parameterisation.\cite{VaTa89}

This has led to the contemporary use of interatomic potentials to be predominantly
in the calculation of the properties of larger non-homogeneous systems, which \textit{cannot}
be modelled using a small periodic cell, and for which ab-initio calculations are 
not computationally feasible. The calculation of the strain and relaxed atomic
positions of large supercells is of crucial importance to the semiconductor science 
community.\cite{LoMa10,Schnor2012} This is because the electronic and optical properties of 
heterostructures are strongly influenced by their strain state.\cite{Reilly89}
Furthermore, computationally cheaper continuum models are able to account for 
neither the atomic-scale variation of composition, nor the atomic-scale reduction
in symmetry which have significant effects on electronic properties.\cite{PrKi98,BeZu05}

We present here a set of potential models which are ideally suited to the study of such structures. 
The valence force field (VFF) model that we use is based on that introduced by Martin.\cite{Ma1970} It 
is well-established how to determine both macroscopic elastic constants and quantities such as 
internal elastic constants and the Kleinman parameter from a given set of VFF parameters. We 
show here that it is possible for diamond and zincblende (ZB) structures to solve the inverse problem, 
namely to calculate VFF parameters for a given material based on known elastic constant and internal strain values. 
The analysis also derives stability criteria for different versions of the VFF model, allowing a simple but 
accurate model for covalent and weakly ionic materials, with additional Coulombic terms required for 
more ionic materials. Having introduced the different models, we then use the results of 
previous \textit{ab-initio} density functional theory calculations to derive and present VFF 
parameter sets for a series of III-V ZB materials. The VFF models presented are 
straightforward to implement in existing atomic simulation packages such as 
LAMMPS\cite{LAMMPS} and GULP,\cite{GULP} thereby allowing the calculation of atomic relaxation 
and strain with a high degree of accuracy, efficiency and physical clarity.
 
Previously, the potential best known for analytic calculation parameters is that of Keating.\cite{Keating66} 
This uses only two VFF parameters, which are determined analytically from the elastic constants $C_{11}$ and $C_{12}$. The model then
describes those two parameters exactly, and captures other elastic 
properties/constants reasonably well. In diamond structure Si, for example, 
the Keating potential will give exact $C_{11}$ and $C_{12}$, and $C_{44}$ with a 1\% error.\cite{Keating66}
Furthermore, while the Keating potential is limited to 
modelling a particular strain regime of a particular crystal phase, it is not, 
due to the analytic expressions for the force constants, limited to any particular 
material. In addition to the accuracy, efficiency and cross-material transferability exhibited
by the Keating potential and others in its class, the simplicity of these models
allows not only for the prediction of the behaviour of large complicated systems, but also 
for its explanation. Because of these advantages, the Keating potential remains
widely used for the calculation of strain and atomistic relaxation 
in large systems, such as semiconductor quantum dots
comprising millions of atoms.\cite{NiRa12,Ziel2012,Zi2013,Ramzi16}

Unfortunately, to describe the elasticity of cubic crystals fully requires more than two elastic constants 
(and more than two force constants), and the Keating potential fares less well for materials other than Si.
For heteropolar materials errors in $C_{44}$ grow with ionicity, and these errors manifest in the inaccurate modelling of systems
where shear strains or internal relaxations are important.\cite{Zi2013,Ziel2012,PrKi98}

The model that we present here, by including details of the inner elasticity of ZB and diamond crystals, 
improves on the accuracy of the Keating model for the description of the elasticity of ZB and diamond structure materials, 
but retains a simple analytic relation between the potential force constants and the elastic properties of the material.
The model possesses the following attractive features: i) it can be immediately applied to any diamond or ZB structure
material for which the required elastic constants are known, with no numerical fitting required; ii) it offers an exact 
description of $C_{11}$, $C_{12}$, $C_{44}$ and the Kleinman parameter $\zeta$, thus providing significantly improved
accuracy over the traditional Keating model, as well as the advantages of improved accuracy and computational efficacy over
more complex potentials; iii) analytic expressions for the force constants allow for clear interpretation and explanation
of results, as well as a-priori prediction of crystal properties other than those by which the potential was parameterised; iv) as noted above,
the simple functional form of the potential is available in most molecular dynamics or crystal energy packages, such as
LAMMPS or GULP (unlike the squared dot products of the Keating model), meaning that anyone with access to these or similar 
packages can use the potential immediately.

In the next section, the elasticity of ZB and diamond structure crystals is described, 
followed in Section III by an outline of the method by which
the force constants of an interatomic potential may be analytically
related to the constants governing the elastic response of any ZB or diamond crystal.
%
We then present in Section~\ref{sec:Non-Coul} the solution of the inverse problem for 
the covalent model, and an investigation of the stability of the covalent model 
for the III-V materials considered, with further details of the analysis 
included in Appendix~\ref{appendix1}. Section~\ref{sec:Conv-Coul}  and \ref{sec:Free-Coul} introduce electrostatic 
interactions into the VFF model, with the values of the effective charges 
determined from the measured optical phonon splitting for each material 
in Sec.~\ref{sec:Conv-Coul} and using the internal elastic constant $E_{11}$ in Sec.~\ref{sec:Free-Coul}. 
In Sec.~\ref{sec:Benchmarking}, the potentials are benchmarked against first principles and experimental results.
Finally, the results are summarised and conclusions presented in Sec.~\ref{sec:Conclusion}.  
%

\section{Theory} \label{sec:theory}
\begin{figure}[h]
   \centering
   \includegraphics[width=0.40\textwidth]{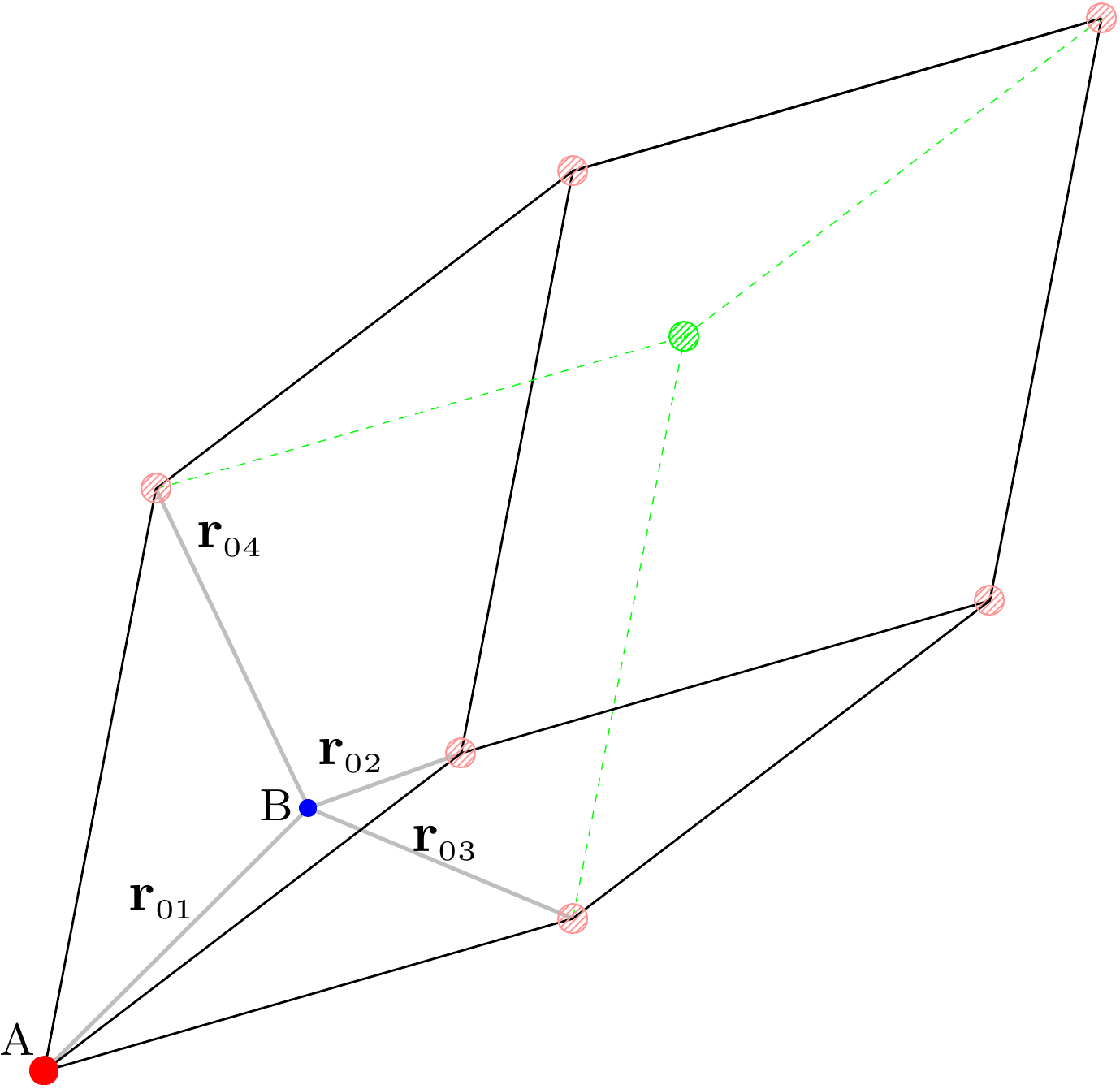}
   \caption{Zincblende primitive cell.}
   \label{fig:cell}
\end{figure}
The primitive unit cell of a ZB or diamond crystal is shown in Fig.~\ref{fig:cell}. The cell
consists of two interpenetrating face centred cubic lattices. The cell can be strained as a whole, and displacements
can also occur between the sublattices, known as \emph{internal strain}.\cite{Born_dynamic,CousinsInner} 
In the harmonic regime macroscopic distortions of the whole
cell are completely specified by the strain tensor $\varepsilon$, and the internal strain between the sublattices is
specified by the internal strain vector $\mathbf{u}$.

The free energy per unit mass per unit volume of a ZB or diamond 
crystal for a general state of (small) macroscopic and internal strain consistent
with its cubic symmetry is given by:
\begin{equation}
\begin{split} \label{eq:MacroMicroU}
 U &= \frac{1}{2}C_{11}\left(\varepsilon_{1}^{2}+\varepsilon_{2}^{2}+\varepsilon_{3}^{2}\right)
 +C_{12}\left(\varepsilon_{1}\varepsilon_{2}+\varepsilon_{1}\varepsilon_{3}+\varepsilon_{2}\varepsilon_{3}\right)\\
 &+\frac{1}{2}C^{0}_{44}\left(\varepsilon_{4}^{2}+\varepsilon_{5}^{2}+\varepsilon_{6}^{2}\right)
 +D_{14}\left(u_{x}\varepsilon_{4}+u_{y}\varepsilon_{5}+u_{z}\varepsilon_{6}\right)\\
 &+\frac{1}{2}E_{11}\left(u_{x}^{2}+u_{y}^{2}+u_{z}^{2}\right) .
\end{split}
\end{equation}
Here the notation of Cousins\cite{CousinsThesis,CousinsInner,CousinsPhysBI} is utilised for the
elastic and inner elastic constants, and we have also above employed Voigt~\cite{Voigt,Nye_Book} notation, 
which, using the symmetry of the strain tensor, makes the convenient contraction of 
indices: 11$\rightarrow$1, 22$\rightarrow$2, 33$\rightarrow$3, 23$\rightarrow$4, 
13$\rightarrow$5, 12$\rightarrow$6. In the above equation $C_{11}$ and $C_{12}$ are the familiar
second-order elastic constants of a cubic crystal, which may be readily obtained from experiment, 
while $C_{44}^{0}$ is the experimentally unobtainable unrelaxed or ``bare'' $C_{44}$ (also known as the 
clamped-ion contribution to the elastic constant $C_{44}$\cite{NiMa85,Miguel_Stress}); 
$C_{44}^{0}$ governs how the crystal responds to shear strains
when the internal strain is set equal to zero. The constant $D_{14}$ accounts for coupling between
internal and macroscopic strain, and the term $E_{11}$ describes the contribution to the free
energy from a pure internal strain. 

$E_{11}$ may be related to the zone-centre transverse optical phonon
frequency and can thus be obtained indirectly from experiment. This relation is given by:\cite{CousinsThesis,NiMa85,VaTa89}
\begin{equation} \label{eq:omegaTO_Bxx}
 E_{\scalebox{0.75}{11}} = 4\mu \omega_{\textrm{\scalebox{0.75}{TO}}}^2 / a_{\textrm{\scalebox{0.75}{0}}}^3 ,
\end{equation}
where $\mu$ is the reduced mass of the anion and cation system, $\omega_{\textrm{\scalebox{0.75}{TO}}}$ is 
the transverse optical phonon frequency at $\Gamma$, and $a_{\textrm{\scalebox{0.75}{0}}}$ is the lattice constant.
The remaining two constants, $C_{44}^{0}$ and $D_{14}$ may be obtained by considering the crystal
energy once it is minimised with respect to internal strain, $\mathbf{u}$. The value of the internal strain
which minimises the free energy is given by:
\begin{equation} \label{eq:txKlein}
 \mathbf{u}^{0} = \left(-\frac{a_0}{4}\zeta\varepsilon_{4},-\frac{a_0}{4}\zeta\varepsilon_{5},-\frac{a_0}{4}\zeta\varepsilon_{6}\right) ,
\end{equation}
where $\zeta$ is Kleinman's internal strain parameter,\cite{Klein62} given by:
\begin{equation} 
\label{eq:disinKleined}
 \zeta = \frac{\sqrt{3}}{r_{0}}\frac{D_{14}}{E_{11}}.
\end{equation}
Though very difficult to perform, especially for more brittle crystals,\cite{CousinsThesis} 
measurements of the Kleinman parameter, $\zeta$, have been
made for a limited number of materials. For example, there are experimental values
of the Kleinman parameter in the literature for Si,\cite{CoGe87} Ge,\cite{CoGe87} 
GaAs,\cite{CousinsGaAsKlein} C,\cite{CoGe89} and InSb.\cite{CoGe91} However, reflecting a general trend
for inner elastic properties, first principles determinations of the Kleinman parameter are abundant for
most group IV or III-V cubic materials.\cite{NiMa85,CaSc15,TaCa19} 

Substituting eq.~(\ref{eq:txKlein}) into eq.~(\ref{eq:MacroMicroU}) then gives the familiar expression for the free
energy which is minimised with respect to internal strain:
\begin{equation}
\begin{split} \label{eq:MacroU}
 U &= \frac{1}{2}C_{11}\left(\varepsilon_{1}^{2}+\varepsilon_{2}^{2}+\varepsilon_{3}^{2}\right)
 +C_{12}\left(\varepsilon_{1}\varepsilon_{2}+\varepsilon_{1}\varepsilon_{3}+\varepsilon_{2}\varepsilon_{3}\right)\\
 &+\frac{1}{2}C_{44}\left(\varepsilon_{4}^{2}+\varepsilon_{5}^{2}+\varepsilon_{6}^{2}\right) ,
\end{split}
\end{equation}
where $C_{44}$ is now the experimentally measurable $C_{44}$, reduced from its unrelaxed value by:
\begin{equation} \label{eq:relax_C44}
 C_{44} = C_{44}^{0} - \frac{D_{14}^{2}}{E_{11}} = C_{44}^{0} - \frac{r_{0}^{2}}{3}\zeta^{2} E_{11}.
\end{equation}

Given the above dependencies amongst the relaxed and unrelaxed elastic constants, then,
if any three independent constants (out of the five: $C_{44}$, $\zeta$, $E_{11}$, $D_{14}$, $C_{44}^{0}$) 
are known, then the remaining two can be obtained indirectly.
Likewise, if any interatomic potential is able to accurately model $C_{11}$, $C_{12}$, and three of these constants,
then the free energy density under any combination of strain or sublattice displacement will be fully described.

To relate these components of the free energy density to the force constants of an interatomic potential,
the interatomic potential is used to express the energy of an arbitrarily deformed primitive diamond or ZB cell. 
This energy is divided by the equilibrium cell volume to obtain the free energy density, and then the VFF energy, 
expressed naturally as a function of the distance between the two atoms in the primitive cell, is cast
in terms of the strain and internal strain:  
\begin{equation}\label{eq:vff_trans}
 U^{\scalebox{0.75}{\textrm{VFF}}}(r_{ij},\theta_{ijk})\Longrightarrow U^{\scalebox{0.75}{\textrm{VFF}}}(\varepsilon, \mathbf u).
\end{equation}
Here we have denoted the primitive cell energy density, expressed in
terms of the VFF force constants as $U^{\scalebox{0.75}{\textrm{VFF}}}$. 
Generally, the expression of the VFF energy in terms of the strain and internal strain
, $U^{\scalebox{0.75}{\textrm{VFF}}}(\varepsilon, \mathbf u)$, will be a very complicated and long function of $\varepsilon$.
However, we are only interested in harmonic elastic properties, so it can 
therefore be expanded in a Taylor series about the equilibrium and 
truncated to second order.

To effect the transformation of eq.~(\ref{eq:vff_trans}), consider Fig.~\ref{fig:cell}.
Keeping the atom at the origin of the cell fixed, the interatomic bond lengths 
can be expressed in terms of the strain and internal strain through the 
transformation with strain of the atomic position vectors:
\begin{equation}
 \begin{aligned}
  \mathbf{r}_{A} &= \mathbf{r}_{A,0} = \left[0,0,0\right] ,\\
  \mathbf{r}_{B} &= \left(I+\varepsilon\right)\mathbf{r}_{B,0}+\mathbf{u}.
 \end{aligned}
\end{equation}
Here, $\mathbf{r}_{A}$ ($\mathbf{r}_{B}$) is the position of the atom labelled A (B) in Fig.~\ref{fig:cell},
with $\mathbf{r}_{A,0}$ ($\mathbf{r}_{B,0}$) being the equilibrium position of this atom, and $I$ is
the $3\times3$ identity matrix.
Substituting these position vectors into the expression for the VFF energy will give the 
energy in terms of the strain, which can then be truncated to second order. 
This procedure has been detailed by Keating.\cite{Keating66} 

Following this 
expansion, direct analytic relations between the elastic and inner elastic constants and
the force constants may be obtained via the derivatives:
\begin{equation}
\begin{aligned}
 &C_{11} = \frac{\partial^{2} U^{\scalebox{0.75}{\textrm{VFF}}}}{\partial\varepsilon_{1}^{2}}    ; &C_{12} = \frac{\partial^{2} U^{\scalebox{0.75}{\textrm{VFF}}}}{\partial\varepsilon_{1}\partial\varepsilon_{2}} \\
 &C_{44}^{0} = \frac{\partial^{2} U^{\scalebox{0.75}{\textrm{VFF}}}}{\partial\varepsilon_{4}^{2}}; &D_{14} = \frac{\partial^{2} U^{\scalebox{0.75}{\textrm{VFF}}}}{\partial u_{x}\partial\varepsilon_{4}} \\
 &E_{11} = \frac{\partial^{2} U^{\scalebox{0.75}{\textrm{VFF}}}}{\partial u_{x}^{2}}             ; &C_{44} = \frac{\partial^{2} U^{\scalebox{0.75}{\textrm{VFF}}}\left(\mathbf{u}=\mathbf{u}_{0}\right)}{\partial\varepsilon_{4}^{2}} \\
 &\zeta = \frac{-4u_{x}^{0}\left(\varepsilon_{4}\right)}{a_{0}\varepsilon_{4}} .
 \end{aligned} \label{Vffelastics}
\end{equation}

In the next section, we present the VFF model with which we model the elastic energy
density described above.

\section{Interatomic Potential} \label{sec:Interatomic Potential}
The VFF with which we describe the elastic properties of diamond and ZB crystals 
was originally introduced by Musgrave and Pople.\cite{MusPop62} We shall follow 
the developments made on this potential by Martin.\cite{Ma1970}
%
Discarding a purportedly unimportant cross angle term, and including terms which account for
the Coulomb interaction between the partially charged ions of a heteropolar crystal, 
Martin gives the form of the potential which will be used in this work. 
For each atom in a ZB crystal Martin's potential is given by: 
\begin{equation} \label{eq:vff}
 \begin{split}
 V_{i} &= \frac{1}{2}\sum_{j\neq i}\frac{1}{2}k_{r}\left(r_{ij}-r_{ij}^{0}\right)^{2} \\
 &+\sum_{j\neq i}\sum_{k\neq i,k>j}\biggl\{\frac{1}{2}k_{\theta}^{i}r_{ij}^{0}r_{ik}^{0}\left(\theta_{ijk}
 -\theta_{ijk}^{0}\right)^{2} \\
 &+k_{r\theta}^{i}\left[r_{ij}^{0}\left(r_{ij}-r_{ij}^{0}\right)
 +r_{ik}^{0}\left(r_{ik}-r_{ik}^{0}\right)\right]\left(\theta_{ijk}-\theta_{ijk}^{0}\right) \\
 &+k_{rr}^{i}\left(r_{ij}-r_{ij}^0\right)\left(r_{ik}-r_{ik}^{0}\right)\biggr\} \\
 &+\frac{1}{2}\sum_{j\neq i}^{\prime} \frac{Z^{*}_{i}Z^{*}_{j}e^{2}}{4\pi\epsilon_{r}\epsilon_{0}r_{ij}} 
 -\frac{1}{2}\sum_{j\neq i}^{nn}\frac{1}{4}\alpha_{M} 
 \frac{Z^{*}_{i}Z^{*}_{j}e^{2}}{4\pi\epsilon_{r}\epsilon_{0}{r^{0}_{ij}}^{2}}\left( r_{ij} - r_{ij}^{0} \right).
 \end{split}
\end{equation}

Here $i$ refers to the central atom being considered, while $j$ and $k$ run over the 4 nearest neighbours for each $i$,
except for the summation $\sum_{j\neq i}^{\prime}$, which runs over the whole crystal. 
This means that in modelling the energy of a ZB primitive cell 8 bond 
lengths and 12 angles will be treated. The half preceeding all two body terms prevents double
counting when summing over $i$, to obtain the energy of the whole crystal from the energy per atom. 
$r_{ij} = \left(\mathbf{r}_{ij}\cdot\mathbf{r}_{ij}\right)^{\frac{1}{2}}$ 
refers to the bond length between atom $i$ and $j$, 
$\theta_{ijk}= \textrm{cos}^{-1}\left(\frac{\mathbf{r}_{ij}\cdot\mathbf{r}_{ik}}{|\mathbf{r}_{ij}||\mathbf{r}_{ik}|}\right)$ 
refers to the angle between the bonds $r_{ij}$
and $r_{ik}$, and $r_{ij}^{0}$ and $\theta_{ijk}^{0}$ denote the equilibrium 
bond lengths and bond angles, respectively.

\begin{figure*}[ht]
   \centering
   \includegraphics[width=0.95\textwidth]{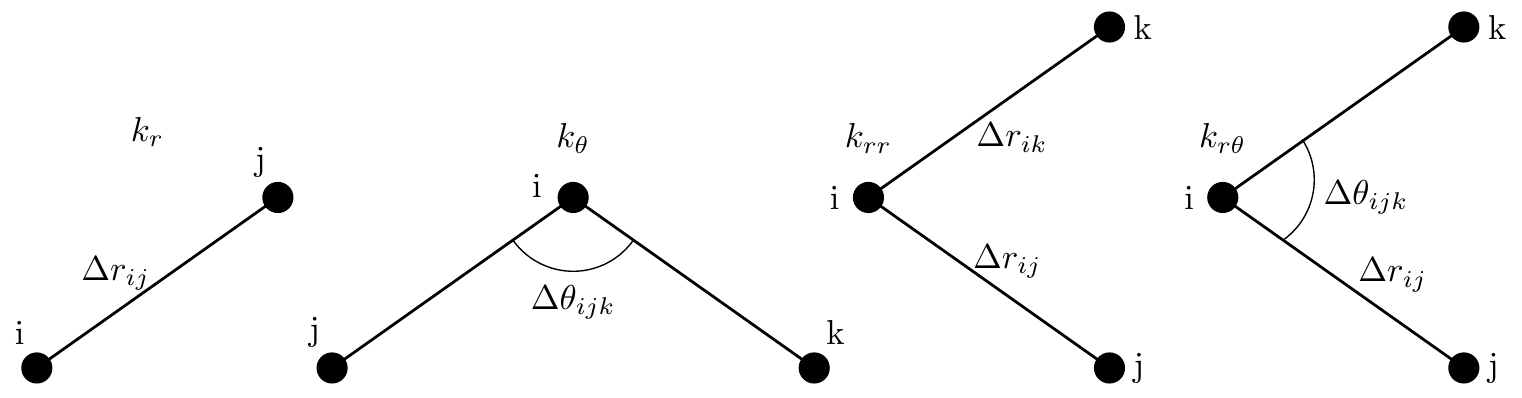}
   \caption{Valence force field interaction terms contributing to eq.~(\ref{eq:vff}). From left to right: bond stretch, $k_{r}$; 
   bond bending, $k_{\theta}$; bond-bond stretching, $k_{rr}$; bond-stretching angle-bending coupling constant, $k_{r\theta}$. }
   \label{fig:terms}
\end{figure*}

The covalent potential terms of eq.~(\ref{eq:vff}) are schematically illustrated in Fig.~\ref{fig:terms}. The term $k_{r}$ captures 
the resistance of any bond to length changes away from the equilibrium length, likewise $k_{\theta}$
describes the harmonic resistance to changes in angle. The term $k^{i}_{rr}$ describes the relation between neighbouring bonds which
share an atom (atom $i$); how one bond will tend to increase in length if another is decreased. 
$k^{i}_{r\theta}$ describes the interaction between the angle between two
bonds, and each of the two bonds; this will, for example, for $k^{i}_{r\theta}>0$, make it energetically favourable for bond lengths to increase when bond angles
decrease. This energetic favourability can be imputed to changes in the $s$-$p$ mixing on the orbitals sitting on the central atom.\cite{KeatingThird} 
The amount by which the energy changes due to this rehybridisation would in principle depend on the species of the central atom; which in turn would imply
different 3-body terms are needed for the cation and the anion, hence the superscript $i$ on these terms. However, Martin justifies 
the exclusion of this effect by emphasising that the potential is being used to study only phenomena in the long-wavelength regime: elastic properties, as well as zone
centre optic and acoustic modes. In this case the force constants for the two atoms in the unit cell always enter the energy and frequency equations together, 
and thus could not be separated, nor would treating them as different result in an improvement in the description of any
of our targeted elastic constants. Anion-centred and cation-centred angular terms are thus treated as the same. 

The last two terms are the terms which account for electrostatic effects, with $Z^{*}$ representing the effective charge
of the ions, and $\alpha_{M}$ denoting the Madelung constant.
The first of these is the screened Coulomb interaction, and the second is a linear repulsive term, given by the linear 
part of the Taylor expansion of the Coulomb energy in the strain, 
necessary to keep the crystal stable at equilibrium, and also to preserve the symmetry of
the elastic constant tensor. The prime symbol over the summation of Coulomb 
interaction indicates it is a long-ranged interaction which must be computed over the whole crystal.

In the work by Martin further approximations and dependencies were applied to the force constants
such that eq.~(\ref{eq:vff}) becomes equivalent to the Keating potential with additional 
Coulombic terms. In this work no dependencies amongst the force constants are imposed, 
and there are thus four force constants and an effective charge 
with which we can describe the elastic properties. For the description of
the elastic energy density given in eq.~(\ref{eq:MacroMicroU}), no advantage can be 
expected from including any further force constants in the VFF model, given 
the arbitrariness in choice of parameter values when fitting six or more 
VFF parameters to the five independent elastic constants.

To obtain the numerical values for the force constants in eq.~(\ref{eq:vff}), the 
potential must be expanded after the manner of Keating, as described in Sec.~\ref{sec:theory}. 
However, this procedure is not straight forward for the Coulomb term; in this case a numerical
Ewald summation must be performed for different strained crystal states to determine the dependence of the Coulomb
energy of the whole crystal on strain. The expansion of our potential in eq.~(\ref{eq:vff}) in terms of strain
and sublattice displacement is given in eq.~(\ref{eq:Vff_strain}) below:
\begin{widetext}
\begin{multline} \label{eq:Vff_strain}
 U = \frac{\sqrt{3}}{2}\left(\frac{1}{12r_{0}}\left(k_{r}+6k_{rr}+12k_{\theta}\right) + \frac{3\alpha_{1}}{8}SC_{0}     \right)
 \left[\varepsilon_{1}^{2}+\varepsilon_{2}^{2} +\varepsilon_{3}^{2}\right]
 + \left(\frac{\sqrt{3}}{12r_{0}}\left(k_{r}+6k_{rr}-6k_{\theta}\right) + \frac{3\sqrt{3}\alpha_{2}}{16}SC_{0} \right)
 \left[\varepsilon_{1}\varepsilon_{2}+\varepsilon_{1}\varepsilon_{3}+\varepsilon_{2}\varepsilon_{3}\right] \\
 + \frac{1}{2}\left(\frac{\sqrt{3}}{12r_{0}}\left(k_{r}-2k_{rr}+4\sqrt{2}k_{r\theta}+2k_{\theta}\right)  +  \frac{3\sqrt{3}\alpha_{2}}{16}SC_{0}      \right)
 \left[\varepsilon_{4}^{2}+\varepsilon_{5}^{2}+\varepsilon_{6}^{2}\right] \\
 +\left(\frac{1}{4r_{0}^{2}}\left(k_{r}-2k_{rr}-2\sqrt{2}k_{r\theta}-4k_{\theta}\right)    +  \frac{9\alpha_{4}}{64r_{0}}SC_{0}  \right)
 \left[u_{x}\varepsilon_{1}+u_{y}\varepsilon_{2}+ u_{z}\varepsilon_{3}\right] \\
 + \frac{1}{2}\left(\frac{\sqrt{3}}{4r_{0}^{3}}\left(k_{r}-2k_{rr}-8\sqrt{2}k_{r\theta}+8k_{\theta}\right)  +\frac{9\sqrt{3}\alpha_{3}}{128r_{0}^{2}}SC_{0}       \right)
 \left[u_{x}^{2}+u_{y}^{2}+u_{z}^{2}\right].
\end{multline}
\end{widetext}
Here we follow Martin\cite{Ma1970} and employ the simplifying notation (in S.I. units) of $S$ and $C_{0}$, where $S$ is the
dimensionless quantity, $\frac{Z^{*2}}{\epsilon_{r}}$, and $C_{0}$ has units of GPa and is given by: $\frac{e^2}{4\pi\epsilon_{0}r_{0}^{4}}$.
The quantities $\alpha_{i}$ are the numerical coefficients obtained by performing an Ewald summation at different strains, and
adding to this the strain dependence of the linear repulsive term, which already contains within it the Madelung constant,
$\alpha_{M}$, which is an Ewald summation at zero-strain. So for example, the result of performing an Ewald summation of these
two electrostatic terms of the crystal for different strain states then yields an electrostatic energy which depends on strain as follows:
\begin{multline}
 E_{el} = \alpha_{1}\left(\varepsilon_{1}^{2}+\varepsilon_{2}^{2}+\varepsilon_{3}^{2}\right)SC_{0} \\ +\alpha_{2}\left(\varepsilon_{1}\varepsilon_{2} + \varepsilon_{1}\varepsilon_{3} 
 + \varepsilon_{2}\varepsilon_{3}\right)SC_{0}+\frac{\alpha_{2}}{2}\left(\varepsilon_{4}^{2}+\varepsilon_{5}^{2}+\varepsilon_{6}^{2}\right)SC_{0} \\
 + \alpha_{3}\left(\tilde{u}_{x}^{2} + \tilde{u}_{y}^{2} + \tilde{u}_{z}^{2}\right) SC_{0}
 + \alpha_{4}\left(\tilde{u}_{x}\varepsilon_{4} + \tilde{u}_{y}\varepsilon_{5} + \tilde{u}_{z}\varepsilon_{6}\right) SC_{0} .
\end{multline}
Above we have utilised the notation $\tilde{u}_{i}=\frac{u_{i}}{a_{0}}$ so that 
the expansion coefficients all have the same units. We see reproduced 
above the result, expected from considerations of symmetry, that the coefficient of the electrostatic energy dependence on biaxial strain is double
that of its dependence on shear strain.\cite{Blackman58,Ma1970,StSa11} The numerical values of these
coefficients are given by:
\begin{equation}
\begin{aligned}
\alpha_{1} &= -0.128411 ,  &\alpha_{2} = -0.417608 ,\\
\alpha_{3} &= -6.53970  ,  &\alpha_{4} = -3.62707 .
\end{aligned} \label{eq.ewald1}
\end{equation} 

When both the nearest neighbour terms and the Coulomb term have been so
expanded in the strain, and the resulting energy density compared with eq.~(\ref{eq:MacroMicroU}), we may use eqs.~(\ref{Vffelastics})
to obtain the following expressions for our VFF force constants and effective charge:
\begin{equation} 
\label{equations}
\begin{aligned}
 C_{11}     &= \frac{\sqrt{3}}{12r_{0}}\left( k_{r}+6k_{rr}+12k_{\theta}\right) + \frac{3\sqrt{3}\alpha_{1}}{8}SC_{0} \,   \\  
 C_{12}     &= \frac{\sqrt{3}}{12r_{0}}\left( k_{r}+6k_{rr}-6k_{\theta}\right)  + \frac{3\sqrt{3}\alpha_{2}}{16}SC_{0} \,  \\
 C_{44}^{0} &= \frac{\sqrt{3}}{3r_{0}}\left(k_{r}-2k_{rr}+4\sqrt{2}k_{r\theta}+2k_{\theta}\right) +  \frac{3\sqrt{3}\alpha_{2}}{16}SC_{0}  \, \\
 D_{14}     &= \frac{1}{4r_{0}^{2}}\left(k_{r}-2k_{rr}-2\sqrt{2}k_{r\theta}-4k_{\theta}\right)    + \frac{9\alpha_{4}}{64r_{0}}SC_{0} \, \\
 E_{11}     &= \frac{\sqrt{3}}{4r_{0}^{3}}\left(k_{r}-2k_{rr}-8\sqrt{2}k_{r\theta}+8k_{\theta}\right) + \frac{9\sqrt{3}\alpha_{3}}{128r_{0}^{2}}SC_{0}  .
\end{aligned} 
\end{equation}
The expressions for the relaxed $C_{44}$ and the Kleinman parameter $\zeta$ may 
be obtained from eqs.~(\ref{eq:disinKleined}) and (\ref{eq:relax_C44}).

In what follows, three different parameterisations of the VFF model are presented and discussed.
Because there are many weakly-polar ZB and non-polar diamond structured materials, and because
the Coulomb interaction is the most computationally expensive to implement, in Sec.~\ref{sec:Non-Coul},
we introduce a computationally efficient covalent VFF in which Coulomb terms are neglected. 
This parameterisation is ideally suited to materials like GaAs. However, as we will show
it is not applicable to materials with ionicity and anisotropy past a certain threshold, such as, for example, InP and InAs.
Therefore, in Sec.~\ref{sec:Conv-Coul}, the Coulomb potential is included 
via the conventional parameterisation based on the optic mode splitting,\cite{Ma1970,StSa11,KiLa96} and the effects of its inclusion
on the stabilty of the model are discussed. It is found that this conventional parameterisation results in a general increase in
accuracy of the potential and restores stability for mildly ionic
materials such as InP and InAs; but that this stabilising effect is not sufficiently large for highly ionic and anisotropic
materials such as cubic GaN and AlN. Thus, in Sec.~\ref{sec:Free-Coul} a non-conventional inclusion of the Coulomb interaction
is presented, whereby the VFF is parameterised along with the force constants from the elastic energy density relations,
ensuring stability with respect to any macroscopic or internal strain, and complete specification of the energy density of
any cubic crystal. 

All numerical quantities determined from the presented numerical relations in the following sections make use of the
elastic and Kleinman parameters calculated in Ref.~\onlinecite{TaCa19}, whilst values for the zone-centre optical phonon
frequencies are taken from elsewhere in the literature, and cited as used.
\subsection{Covalent (non-Coulombic) VFF} \label{sec:Non-Coul}
To efficiently model non-polar crystals, we may set $S$, in eqs.~(\ref{equations}), to 0.
Then, using eqs.~(\ref{eq:relax_C44}) and (\ref{eq:disinKleined}), we obtain the following
simplified expressions for $C_{44}$ and $\zeta$:

\begin{eqnarray}
 C_{44} &=& \frac{3\sqrt{3}}{2r_{0}}\frac{k_{r}k_{\theta}-2k_{rr}k_{\theta}-4k_{r\theta}^{2}}
{k_{r}-2k_{rr}-8\sqrt{2}k_{r\theta}+8k_{\theta}} \, \label{eq:C44} , \\
 \zeta &=& \frac{k_{r}-2k_{rr}-2\sqrt{2}k_{r\theta}-4k_{\theta}}{k_{r}-2k_{rr}-8\sqrt{2}k_{r\theta}+8k_{\theta}} \, \label{eq:Klein} .
\end{eqnarray}

We can invert these two equations, along with the expressions for $C_{11}$ and $C_{12}$ in eq.~(\ref{equations}). Taking 
care to eliminate the extraneous root that comes from a quadratic equation derived from $C_{44}$ and $\zeta$ (see Appendix in Sec.~\ref{appendix1}), 
we obtain direct expressions for the force constants in terms of the second order elastic constants and the Kleinman parameter. 
These read:

\begin{widetext}
\begin{eqnarray}
 k_{\theta} &=& \frac{2\left(C_{\scalebox{0.75}{11}}-C_{\scalebox{0.75}{12}}\right)r_{0}}{3\sqrt{3}} \, \label{force1} , \\
 k_{r} &=& \frac{r_{\scalebox{0.75}{0}}\left[C_{\scalebox{0.75}{11}}\left(2+2\zeta+5\zeta^2\right) + 
 C_{\scalebox{0.75}{12}}\left(1-8\zeta-2\zeta^2\right)+ 3C_{\scalebox{0.75}{44}}\left(1-4\zeta\right)\right]}{\sqrt{3}\left(1-\zeta\right)^{2}} \, \label{force2} , \\
 k_{rr} &=& \frac{r_{\scalebox{0.75}{0}}\left[C_{\scalebox{0.75}{11}}\left(2 - 10\zeta - \zeta^2\right) 
 + C_{\scalebox{0.75}{12}}\left(7-8\zeta+10\zeta^2\right)-3C_{\scalebox{0.75}{44}}\left(1-4\zeta\right)\right]}{6\sqrt{3}\left(1-\zeta\right)^{2}} \, \label{force3} , \\
 k_{r\theta} &=& \frac{r_{\scalebox{0.75}{0}}}{3}\sqrt{\frac{2}{3}}\frac{\left(C_{\scalebox{0.75}{11}}-C_{\scalebox{0.75}{12}}\right)\left(1+2\zeta\right)-3C_{\scalebox{0.75}{44}}}{\zeta-1}  \, \label{force4} .
\end{eqnarray}
\end{widetext}

Having this one-to-one analytic relation between the force constants and the elastic constants has several advantages. 
Like the Keating model we have direct expressions for the force constants with no numerical fitting procedures required. 
Thus, unlike potentials for which a numerical fitting is required, and a new fitting is needed for 
each material, eqs.~(\ref{force1}) to (\ref{force4}) ideally represent a VFF for any ZB
or diamond structure material: once the elastic constants and the Kleinman parameter are known,
so too are the force constants. Unlike the Keating potential which describes 
exactly only the elastic constants $C_{11}$ and $C_{12}$, with significant errors often found
for $C_{44}$ and $\zeta$, 
the above relations ensure that these properties
are reproduced exactly. 

With respect to other more sophisticated potentials, this parameterisation
of the VFF model offers all the advantages of simplicity, efficiency and 
clarity of the much used Keating model. The model even offers greater accuracy, 
in the regime for which it is parameterised, when compared to more complex potentials.
\begin{table}[t] 
\centering  
\begin{tabular}{|c|c|c|}
\hline
         & $E_{11}$~(GPa~\AA$^{-2}$)  &  $\omega_{\scalebox{0.5}{TO}}$~(cm$^{-1}$) \\ \hline
  AlN    & -210.9                   &  n/a                                     \\ \hline
  AlP    & 12.06                    &  241 (454\textsuperscript{a}, -46\%)     \\ \hline 
 AlAs    & 11.09                    &  209 (360\textsuperscript{b}, -42\%)     \\ \hline
 AlSb    & 12.45                    &  238 (318\textsuperscript{c}, -25\%)     \\ \hline
  GaN    & -132.9                   &  n/a                                     \\ \hline
  GaP    & 22.45                    &  269 (366\textsuperscript{d}, -27\%)     \\ \hline
 GaAs    & 16.52                    &  189 (273\textsuperscript{d}, -30\%)     \\ \hline
 GaSb    & 13.40                    &  173 (231\textsuperscript{c}, -25\%)     \\ \hline
  InN    & -282.0                   &  n/a                                     \\ \hline
  InP    & -2.62                    &  n/a                                     \\ \hline
 InAs    & -0.3                     &  n/a                                     \\ \hline
 InSb    & 4.28                     &  93 (185\textsuperscript{c}, -49\%)      \\ \hline
\end{tabular}
\caption{Covalent VFF model prediction of $E_{11}$ via eq.~(\ref{eq:bxx_eps}), and zone-centre 
transverse-optical phonon frequency, $\omega_{\textrm{\tiny{TO}}}$, via eq.~(\ref{eq:omegaTO_Bxx}). 
Experimental values and percentage differences between these and predicted values, are given in brackets. 
a= Ref.~\onlinecite{BeerJack68}; b=Ref.~\onlinecite{AzSo95}; 
c=Ref.~\onlinecite{Yu2010}; d=Ref.~\onlinecite{Moor66}.  } \label{tab:Bxx} 
\end{table}

In addition, these simple expressions make the explanation of different trends in elastic properties 
in terms of the force constants a straight forward procedure. 
For example, eqs.(\ref{equations}) and (\ref{force1}-\ref{force4}) may be used to obtain
expressions for the other elastic constants, $C_{44}^{0}$, $E_{11}$ and $D_{14}$.
These expressions may then be used to predict quantities on which the model has not been parameterised 
to ascertain the suitability of the potential for the different materials.

One such prediction is the value of the inner elastic constant $E_{11}$, which 
can be related to the experimental transverse optical phonon mode at the $\Gamma$ point.
While the potential is not aimed at the accurate description of dynamical properties, such quantities
will nevertheless give an indication of whether or not the energetics of, for example, the internal 
strain, are reasonable.
The quantity $E_{11}$ is related to the frequency of the transverse optical phonon mode at $\Gamma$ for 
ZB structures by eq.~(\ref{eq:omegaTO_Bxx}).
From eqs.~(\ref{equations}) and (\ref{force1}) to (\ref{force4}), the following relation
between $E_{11}$ and the known elastic properties is derived:
\begin{equation} \label{eq:bxx_eps}
 E_{11} = \frac{16\left(C_{11}-C_{12}-C_{44}\right)}{\left(1-\zeta\right)^2a_{\textrm{\scalebox{0.75}{0}}}^{2}}.
\end{equation}
A negative $E_{11}$ would lead to two undesirable results: imaginary $\omega_{\textrm{\scalebox{0.75}{TO}}}$ (cf eq.~(\ref{eq:omegaTO_Bxx})), 
and worse, the scenario that the energy density has a stationary point
which is a maximum rather than a minimum in the internal strain; i.e. that the crystal
is unstable with respect to internal strain. This latter consequence 
invalidates the basis of the whole procedure by which the relaxed
elastic constants are derived, wherein the assumption is made that the energy 
is being \emph{minimised} with respect to the internal strain. 
 
\begin{table}[t] 
\centering  
\begin{tabular}{|c|c|c|c|c|c|}
\hline
      & $k_{r}$                           & $k_{\theta}$                    & $k_{rr}$                         & $k_{r\theta}$                                               & $S$ \\ \hline
units & eV~\AA{}$^{\scalebox{0.75}{-2}}$  & eV~rad$^{\scalebox{0.75}{-2}}$  & eV~\AA{}$^{\scalebox{0.75}{-2}}$ & eV~\AA{}$^{\scalebox{0.75}{-1}}$rad$^{\scalebox{0.75}{-1}}$ &     \\ \hline
AlP   & 5.505                             & 0.401                           & 0.640                            & 0.453                                                       & 0.000\\ \hline
AlAs & 4.962                              & 0.361                           & 0.521                           & 0.391                            & 0.000\\ \hline
AlSb & 4.557        & 0.294               & 0.320  & 0.249 & 0.000 \\ \hline
GaP  & 6.237        & 0.464               & 0.455  & 0.421 & 0.000\\ \hline
GaAs & 5.292        & 0.397               & 0.396  & 0.364 & 0.000\\ \hline
GaSb & 4.542        & 0.319               & 0.264  & 0.258 & 0.000\\ \hline
InSb & 3.194        & 0.218               & 0.362  & 0.248 & 0.000\\ \hline
\end{tabular}
\caption{Force constant values for the covalent VFF model for selected III-V semiconductors.} \label{tab:force_consts}
\end{table} 
 
An inspection of the terms in the numerator of eq.~(\ref{eq:bxx_eps}) reveals that only those crystals for which $C_{11}-C_{12} > C_{44}$,
or:
\begin{equation}
\label{eq:aniso_condition}
 A =  \frac{2C_{44}}{C_{11}-C_{12}} < 2,
\end{equation}
where $A$ is the anisotropy parameter,\cite{Kittel_2nd} are stable against sublattice displacements. 
Furthermore, we note that this result is not restricted to our particular potential form, but holds also for similar
covalent potentials with Keating-style coordinates (i.e. a potential which is a function of dot products of bond vectors), and those with additional angle-angle coupling terms such 
as those in Refs.~\onlinecite{MusPop62,TuLu72,VaTa89,CousinsPhysBII,StSa11}. Thus, we may say that no nearest-neighbour
VFF model can simultaneously describe $C_{11}$, $C_{12}$, $C_{44}$ and $\zeta$ for 
crystals with $A >2$. 

Relations of this kind may also be used in guiding numerical fittings away from dead ends, 
with an appropriate choice of fitting weights. For example, eq.~(\ref{eq:bxx_eps}) presents an upper
limit on the accuracy with which the components of the elasticity of a ZB or diamond structured crystal can be
simultaneously described using any nearest neighbour VFF model. The relation shows that, 
for example, in the work of Steiger \textit{et al.},\cite{StSa11} if equal weights in the numerical fitting were given to 
$C_{11}$, $C_{12}$, $C_{44}$, $E_{11}$ and $\zeta$, then it would not be possible to simultaneously minimise the residuals, 
and the fitting would go on forever.

The values of $E_{11}$ predicted from eq.~(\ref{eq:bxx_eps}), and the transverse optical 
phonon frequencies, $\omega_{\textrm{\scalebox{0.75}{TO}}}$, corresponding to these are shown in Table~ \ref{tab:Bxx}.
Table~\ref{tab:Bxx} shows that, with negative predicted values for $E_{11}$ and imaginary $\omega_{\textrm{\scalebox{0.75}{TO}}}$,
the potential is not suitable for the highly ionic cubic III-N or any of the indium containing
III-Vs other than InSb. Simulations of these crystals with negative $E_{11}$ using this potential 
are then unstable with respect to internal displacements.

Table~\ref{tab:Bxx} and the condition defined in eq.~(\ref{eq:aniso_condition}) 
demonstrate that the VFF potential (eq.~(\ref{eq:vff})) parameterised via
eqs.~(\ref{force1}-\ref{force4}), is suitable for neither the structural relaxation nor the dynamics 
of materials for which $A>2$, whilst for materials with $A<2$ the potential
describes the parameters of the structural relaxation very well ($C_{ij}$ and $\zeta$), 
but does not accurately describe the $\Gamma$ point optical phonons. These results have been
further corroborated by actual structural relaxations, where materials with $A<2$ 
relax to the correct equilibrium state and respond correctly to different applied strains.

The force constants for selected III-V materials whose structural relaxation is suitably described by 
the covalent VFF model are given in Table~\ref{tab:force_consts}.

From the fact that the inequality of eq.~(\ref{eq:aniso_condition}) tends to be most
strongly violated by the more ionic compounds, we can infer that the Coulomb interaction
plays an important role in stabilising heteropolar crystals, and that neglecting it
is not justified. We therefore include the Coulomb interaction in the next subsection, using the conventional parameterisation
based on the splitting in zone-centre transverse and longitudinal optical phonon mode frequencies.

\subsection{Conventional inclusion of Coulomb interaction} \label{sec:Conv-Coul}
\begin{table}[t] 
\centering  
\begin{tabular}{|c|c|c|c|c|c|}
\hline
      & $k_{r}$                           & $k_{\theta}$                    & $k_{rr}$                         & $k_{r\theta}$                                               &  $S$   \\ \hline
units & eV~\AA{}$^{\scalebox{0.75}{-2}}$  & eV~rad$^{\scalebox{0.75}{-2}}$  & eV~\AA{}$^{\scalebox{0.75}{-2}}$ & eV~\AA{}$^{\scalebox{0.75}{-1}}$rad$^{\scalebox{0.75}{-1}}$ &        \\ \hline
AlP   & 7.017 &   0.392 &   0.450 &   0.333   & 0.689  \\ \hline
AlAs  & 6.880  &   0.349 &   0.279 &   0.240  & 0.593  \\ \hline
AlSb  & 5.550  &   0.289 &   0.192 &   0.173  & 0.373  \\ \hline
GaP   & 7.841  &   0.453 &   0.263 &   0.287  & 0.510  \\ \hline
GaAs  & 6.520  &   0.389 &   0.250 &   0.261  & 0.448  \\ \hline
GaSb  & 5.068  &   0.315 &   0.199 &   0.215  & 0.222  \\ \hline
InN   & 10.513 &   0.272 &   0.862 &   0.425  & 1.996  \\ \hline
InP   & 5.892  &   0.276 &   0.366 &   0.262  & 0.609  \\ \hline
InAs  & 5.031  &   0.243 &   0.325 &   0.227  & 0.551  \\ \hline
InSb  & 4.272  &   0.213 &   0.215 &   0.172  & 0.384  \\ \hline
\end{tabular}
\caption{Force constant values for selected III-V semiconductors using the Coulombic VFF model fitted
to optical phonon frequency splitting.} \label{tab:ConvCoulforce_consts}
\end{table}

Conventionally\cite{Ma1970,KiLa96,StSa11,GrNe01} the effective charge parameter, $S$, in a VFF potential is determined
from the splitting between the optic mode frequencies at the $\Gamma$ point. This relation is
given in eq.~(\ref{eq:S}) below:
\begin{equation} \label{eq:S}
 S = \frac{Z^{*2}}{\epsilon_{r}} = \left(\frac{\Omega}{4\pi e^{2}}\right)\mu\epsilon_{0}\left(\omega_{\textrm{\tiny{LO}}}^{2}-\omega_{\textrm{\tiny{TO}}}^{2}\right).
\end{equation}
Here $Z^{*}$ is the effective charge, $\epsilon_{r}$ is, in this relation, the high frequency dielectric constant of the material in question, $\Omega$ is
the volume of the primitive cell, $e$ is the electronic charge, $\omega_{\textrm{\tiny{LO}}}$ and $\omega_{\textrm{\tiny{TO}}}$ are the longitudinal and transverse
optical phonon frequencies, respectively, and $\mu$ is the reduced mass of the anion and cation atoms.
With this value for $S$, we may solve eqs.~(\ref{equations}) 
in a similar manner as before to obtain the following expressions 
for the force constants:
\begin{widetext}
\begin{eqnarray}
 k_{\theta} &=& \frac{2\left(C_{11}-C_{12}+\frac{3\sqrt{3}}{8}\left(2\alpha_{2}-\alpha_{1}\right)SC_{0}\right)r_{0}}{3\sqrt{3}} \, \label{Cforce1} , \\
 k_{r} &=& \frac{r_{0}\left[C_{11}\left(2+2\zeta+5\zeta^2\right) + 
 C_{12}\left(1-8\zeta-2\zeta^2\right)+ 3C_{44}\left(1-4\zeta\right) + SC_{0}\left(a_{1} + a_{2}\zeta + a_{3}\zeta^{2}\right)\right]} 
 {\sqrt{3}\left(1-\zeta\right)^{2}} \,  , \\   \label{Cforce2}
 k_{rr} &=& \frac{r_{0}\left[C_{11}\left(2 - 10\zeta - \zeta^2\right) 
 + C_{12}\left(7-8\zeta+10\zeta^2\right)-3C_{44}\left(1-4\zeta\right) + SC_{0}\left(a_{4}+a_{5}\zeta+a_{6}\zeta^{2}\right)\right]}
 {6\sqrt{3}\left(1-\zeta\right)^{2}} \, \label{Cforce3} , \\
 k_{r\theta} &=& \frac{r_{0}}{3}\sqrt{\frac{2}{3}}\frac{\left(C_{11}-C_{12}\right)\left(1+2\zeta\right)-3C_{44}+SC_{0}\left(a_{7}+a_{8}\zeta\right)}
 {\zeta-1}  \, \label{Cforce4} .
\end{eqnarray}
\end{widetext}
\begin{table*}[t] 
\centering  
\begin{tabular}{|c|c|c|c|c|c|c|c|}
\hline
 & $C^{\prime}$-$C_{44}$ (GPa)  &  $S$                       & 0.136 $C_{0}S$ (GPa) & $E_{11}$ GPa \AA$^{\scalebox{0.75}{-2}}$ & $\omega_{\scalebox{0.5}{TO}}$~cm$^{-1}$             &  $\omega_{\scalebox{0.5}{LO}}$~cm$^{-1}$   &  $Z^{*}$    \\ \hline
  AlN    & -53.48               & 1.5454\textsuperscript{a}  &  37.91               &  -61.40 & n/a                                                 &  n/a                                       &  2.73~(2.70\textsuperscript{j})   \\ \hline
  AlP    &   4.00               & 0.6888\textsuperscript{b}  &   6.84               &   32.22 & 342~(454\textsuperscript{b},25\%)                   &  390~(491\textsuperscript{b},21\%)         &  2.28~(-)  \\ \hline
 AlAs    &   4.06               & 0.5931\textsuperscript{c}  &   5.05               &   24.91 & 313~(360\textsuperscript{c},13\%)                   &  361~(402\textsuperscript{c},10\%)         &  2.21~(2.17\textsuperscript{k})  \\ \hline
 AlSb    &   5.03               & 0.3728\textsuperscript{d}  &   2.26               &   18.07 & 287~(323\textsuperscript{d},11\%)                   &  310~(344\textsuperscript{d},10\%)         &  1.95~(1.91\textsuperscript{k}) \\ \hline
  GaN    & -31.31               & 1.3373\textsuperscript{e}  &  29.23               &   -8.83 & n/a                                                 &  n/a                                       &  2.55~(2.65\textsuperscript{l})  \\ \hline
  GaP    &   9.11               & 0.5098\textsuperscript{d}  &   5.11               &   35.03 & 336~(366\textsuperscript{d},8 \%)                   &  376~(403\textsuperscript{d},7 \%)         &  2.16~(2.03\textsuperscript{m}) \\ \hline
 GaAs    &   7.41               & 0.4476\textsuperscript{d}  &   3.81               &   25.01 & 232~(273\textsuperscript{d},15\%)                   &  259~(296\textsuperscript{d},13\%)         &  2.20~(2.19\textsuperscript{n}) \\ \hline
 GaSb    &   6.38               & 0.2224\textsuperscript{f}  &   1.38               &   16.33 & 191~(231\textsuperscript{f},17\%)                   &  202~(240\textsuperscript{f},16\%)         &  1.79~(1.73\textsuperscript{k})  \\ \hline
  InN    & -28.01               & 1.9960\textsuperscript{g}  &  28.64               &    6.37 & 164~(478\textsuperscript{g},66\%)                   &  529~(694\textsuperscript{g},24\%)         &  4.09~(3.02\textsuperscript{j})    \\ \hline
  InP    &  -0.69               & 0.609 \textsuperscript{h}  &   5.08               &   14.29 & 226~(307\textsuperscript{h},26\%)                   &  294~(343\textsuperscript{h},14\%)         &  2.58~(2.38\textsuperscript{m}) \\ \hline
 InAs    &  -0.10               & 0.5507\textsuperscript{i}  &   3.50               &   11.10 & 154~(217\textsuperscript{i},29\%)                   &  185~(240\textsuperscript{i},23\%)         &  2.61~(-) \\ \hline
 InSb    &   1.55               & 0.3839\textsuperscript{i}  &   1.84               &    9.54 & 139~(180\textsuperscript{i},23\%)                   &  155~(192\textsuperscript{i},20\%)         &  2.45~(-)  \\ \hline 
\end{tabular}
\caption{Properties relevant to, and predicted from, the conventionally parameterised Coulombic VFF. 
First four columns are related to eq.~(\ref{eq:E11_ConvCoul}) and the predicted value of the internal elastic constant, $E_{11}$. 
$C^{\prime}$ = $C_{11}-C_{12}$, and $C_{44}$ are obtained from Ref.~\onlinecite{TaCa19}, $S$
is determined using eq.~(\ref{eq:S}) with experimental phonon frequencies, and $C_{0}$ is the quantity $\frac{e^2}{4\pi\epsilon_{0}r_{0}^{4}}$. 
The $\omega_{\textrm{\tiny{TO}}}$ and $\omega_{\textrm{\tiny{LO}}}$ columns compare VFF-predicted phonon 
frequencies with experiment and the $Z^{*}$ column gives effective charges obtained from 
experiment via eq.~(\ref{eq:S}), using values for $\varepsilon_{r}$ from Ref.~\onlinecite{MadelHB}, and gives in brackets, where available, 
ab-initio values.
%
Superscripts a-k indicate where experimental values of $\omega_{\textrm{\tiny{TO}}}$ and $\omega_{\textrm{\tiny{LO}}}$, or theoretical values of 
$Z^{*}$ were obtained:
a=Ref.~\onlinecite{KiLa96}; b=Ref.~\onlinecite{BeerJack68}; c=Ref.~\onlinecite{AzSo95}; 
d=Ref.~\onlinecite{Moor66}; e=Ref.~\onlinecite{KaWa98}; f=Ref.~\onlinecite{HaHe62}; g=Ref.~\onlinecite{PropIIINs};
h=Ref.~\onlinecite{MadelHB}; i=Ref.~\onlinecite{LoYu05}; j=Ref.~\onlinecite{BeFi97}; k=Ref.~\onlinecite{GiGi91}; 
l=Ref.~\onlinecite{KaWa98}; m=Ref.~\onlinecite{SeBi95}; n=Ref.~\onlinecite{WaVa07}. } \label{tab:Conv_E11}
\end{table*}
Here the $a_{i}$ denote combinations of Ewald summation terms:
\begin{eqnarray*}
a_{1} &=& -\frac{12\sqrt{3}}{128}\left(8\alpha_{1}+8\alpha_{2}+3\alpha_{4}\right) , \\
a_{2} &=& -\frac{6\sqrt{3}}{128}\left(16\alpha_{1}-80\alpha_{2}-3\alpha_{3}\right)  , \\
a_{3} &=& -\frac{3\sqrt{3}}{128}\left(80\alpha_{1}-16\alpha_{2}-3\alpha_{3}+24\alpha_{4}\right)  , \\
a_{4} &=& \frac{12\sqrt{3}}{128}\left(-8\alpha_{1}   - 8\alpha_{2}  + 3\alpha_{4}\right) ,
\end{eqnarray*}
\begin{eqnarray*}
a_{5} &=& \frac{6\sqrt{3}}{128} \left( 80\alpha_{1}  - 16\alpha_{2} - 3\alpha_{3}\right) , \\ 
a_{6} &=& \frac{3\sqrt{3}}{128} \left( 16\alpha_{1}  - 80\alpha_{2} - 3\alpha_{3} + 24\alpha_{4}\right) , \\
a_{7} &=& -\frac{6\sqrt{3}}{128}\left(8\alpha_{1} - 16\alpha_{2} + 3\alpha_{4}\right) , \\
a_{8} &=& -\frac{3\sqrt{3}}{128}\left(32\alpha_{1} - 16\alpha_{2} - 3\alpha_{3} + 6\alpha_{4}\right) .
\end{eqnarray*} \label{eq.ewald}
Note that eqs.~(\ref{Cforce1}-\ref{Cforce4}) are identical to the covalent equations (eqs.~(\ref{force1}-\ref{force4})) apart from
the electrostatic addition; the non-Coulombic case can be recovered by setting $S=0$.
Force constants and effective charge parameters for selected III-V materials obtained from eqs.~(\ref{Cforce1}-\ref{Cforce4}) are
given in Table~\ref{tab:ConvCoulforce_consts}.

Using these new force constant expressions, the inner elastic constant, $E_{11}$, predicted by
the model is given by the relation:
\begin{equation}
\label{eq:E11_ConvCoul}
 E_{11} =  \frac{16\left(C_{11}-C_{12}-C_{44}+0.135645~C_{0}S\right)}{\left(1-\zeta\right)^2a_{\textrm{\scalebox{0.75}{0}}}^{2}},
\end{equation} 
where the numerical factor 0.135645 results from the sum:$\frac{3\sqrt{3}}{8}\left(-\alpha_{1}+\alpha_{2}+\frac{\alpha_3}{16}-\frac{\alpha_{4}}{4}\right)$.
From this equation the stabilising effect of the Coulomb interaction is apparent: the larger the product
$SC_{0}$, the less strict need be the inequality $C_{11}-C_{12} >C_{44}$ to maintain stability. Thus,
materials with an anisotropy parameter $A>2$, which are unstable in the purely covalent model, can be stabilised
by the inclusion of Coulomb effects. Table~\ref{tab:Conv_E11} illustrates this for the parameterisation used here, where 
the calculated value of $E_{11}$ is given for the III-V materials that we consider.

Table~\ref{tab:Conv_E11} shows that while many materials unstable in the non-Coulombic case
have become stable, the Coulomb interaction derived from eq.~(\ref{eq:S}) is not sufficeint to stabilise the highly ionic 
cubic III-N materials AlN and GaN. Furthermore, we note that while InN is stable whilst utilising 
the optical phonon splitting of Kim \textit{et al.},\cite{KiLa96} using other results (e.g. from Ref.~\onlinecite{KaKa00}) for
$\omega_{\textrm{\tiny{TO}}}$ and $\omega_{\textrm{\tiny{LO}}}$ will yield a smaller value for $S$ and an unstable crystal.

Nevertheless, the values of $\omega_{\textrm{\tiny{TO}}}$ derived using eq.~(\ref{eq:omegaTO_Bxx}) 
and presented in Table~\ref{tab:Conv_E11} reveal a universal 
reduction in the error, when compared with the non-Coulombic results presented
in Table~\ref{tab:Bxx}.
Furthermore, with the addition of the Coulomb interaction, the qualitative description of the zone-centre
optical phonons is greatly improved, with $\omega_{\textrm{\tiny{TO}}}$ and $\omega_{\textrm{\tiny{LO}}}$ no longer
degenerate. The values of $\omega_{\textrm{\tiny{TO}}}$ and $\omega_{\textrm{\tiny{LO}}}$ predicted using the VFF described
by eqns.~(\ref{eq:S}) and~(\ref{Cforce1}-\ref{Cforce2}) are given in Table~\ref{tab:Conv_E11}. In addition, we 
see from Table~\ref{tab:Conv_E11} that the effective charge parameter $S$, obtained from experiment via eq.~(\ref{eq:S}),
produces Born effective charges, $Z^{*}$, which are in good agreement with those determined from first-principles calculations.

However, given our aim is to completely describe the elastic energy of any ZB or diamond
structure material, the instabilities found for AlN and GaN lead to the
conclusion that for this VFF model, the conventional Coulomb parameterisation is not appropriate
when modelling highly ionic materials.
Other approaches to the parameterisation of the effective charge exist in the literature:
for example, Grosse and Neugebauer\cite{GrNe01} used the difference in the total energies of ZB and wurtzite phases
of the III-N materials, AlN, GaN, and InN to determine the effective charge; and Barret and Wang\cite{BaWa15} introduced a model
where the atomic charge is separated from the Born effective charge, and both are utilised in a double charge model for 
the accurate treatment of the lattice dynamics of surfaces. However, in
both of these methods, the charge parameter $S$ produced is smaller than that obtained using eq.~(\ref{eq:S}).

Therefore, in the next section, we seek a more direct means of ensuring that the VFF model
correctly describes the dependence of the energy of the crystal on the internal strain.
This involves breaking with the conventional parameterisation of the effective charge,
and setting the parameter $S$ such that the inner elastic constant $E_{11}$ is exactly reproduced.
With this parameterisation, 
the elastic energy density and internal strain of a ZB or diamond crystal will then be well described
by the VFF for any combination of macroscopic and internal strain.
\begin{table}[t] 
\centering  
\begin{tabular}{|c|c|c|c|c|c|}
\hline
      & $k_{r}$                           & $k_{\theta}$                    & $k_{rr}$                         & $k_{r\theta}$ &  $S$   \\ \hline
units & eV~$\textrm{\AA{}}^{\scalebox{0.75}{-2}}$  & eV~rad$^{\scalebox{0.75}{-2}}$  & eV~$\textrm{\AA{}}^{\scalebox{0.75}{-2}}$ & eV~$\textrm{\AA{}}^{\scalebox{0.75}{-1}}$rad$^{\scalebox{0.75}{-1}}$ &  \\ \hline
  AlN  &  23.52  &  0.506  & -0.024  &   0.517 & 3.378    \\ \hline
  AlP  &   9.30  &  0.379  &  0.162  &   0.361 & 1.046    \\ \hline
 AlAs  &   8.00  &  0.343  &  0.139  &   0.371 & 0.9387   \\ \hline
 AlSb  &   6.42  &  0.284  &  0.081  &   0.283 & 0.6991   \\ \hline
  GaN  &  19.17  &  0.536  &  0.239  &   0.696 & 2.45786  \\ \hline
  GaP  &   8.67  &  0.447  &  0.165  &   0.514 & 0.7721  \\ \hline
 GaAs  &   7.90  &  0.379  &  0.087  &   0.357 & 0.9490 \\ \hline
 GaSb  &   6.43  &  0.307  &  0.032  &   0.275 & 0.8052 \\ \hline
  InN  &  14.75  &  0.263  &  0.220  &   0.467 & 2.3266  \\ \hline
  InP  &   7.71  &  0.269  &  0.115  &   0.356 & 1.04947 \\ \hline
 InAs  &   6.85  &  0.235  &  0.077  &   0.264 & 1.0794 \\ \hline
 InSb  &   5.49  &  0.208  &  0.049  &   0.246 & 0.8252 \\ \hline
\end{tabular}
\caption{Force constant values determined using the Coulombic VFF model 
with effective charges 
determined by elastic and inner elastic properties.} \label{tab:FreeCoulforce_consts}
\end{table}
\subsection{Free parameterisation of effective charge} \label{sec:Free-Coul}

In order to guarantee that the elastic energy density is completely described by
our VFF model we include the inner elastic constant $E_{11}$ in the fitting, and solve for $S$ such
that the correct, positive value is reproduced. 
Thus the interaction parameters $k_{r}$, $k_{\theta}$, $k_{rr}$, $k_{r\theta}$, and $S$ are 
obtained from the known elastic constants $C_{11}$, $C_{12}$, $C_{44}$, $\zeta$ and $E_{11}$.
This ensures not only that the crystal will be stable against shear and internal strains,
since we are fitting directly to a positive $E_{11}$, but also that the dependence of the free energy
of any diamond or ZB crystal on any combination of macroscopic or internal strain, will be described completely. 
Allowing $S$ to be set in this way is justified because there is in any case some degree of arbitrariness in 
the choice of the  \emph{effective} charge, 
given delocalisation and screening effects present in the crystal.

To achieve this parameterisation we make use of eq.~(\ref{eq:E11_ConvCoul}), which gives 
the value of $E_{11}$ in terms of $C_{11}$, $C_{12}$, $C_{44}$, $\zeta$, and $S$. We now solve this equation for $S$,
to obtain the following expression:
\begin{equation} \label{eq:S_E11}
 S= \frac{E_{11}\left(1-\zeta\right)^{2}a_{0}^{2}-16\left(C_{11}-C_{12}-C_{44}\right)}{6\sqrt{3}C_{0}\left(-\alpha_{1}+\alpha_{2}+\alpha_{3}/16-\alpha_{4}/4\right)} .
\end{equation}
Substituting the value for $S$ thus obtained into eqs.~(\ref{Cforce1}-\ref{Cforce4}) yields the required
potential.

\begin{table}[t] 
\centering  
\begin{tabular}{|c|c|c|}
\hline
         & $\omega_{\scalebox{0.5}{TO}}$~(cm$^{-1}$)  &  $\omega_{\scalebox{0.5}{LO}}$~(cm$^{-1}$) \\ \hline
  AlN    & 654~(654\textsuperscript{a},0\%)                             &  1145(908\textsuperscript{a},-26\%)                                     \\ \hline
  AlP    & 454~(454,0\%)                             &  542 (491, -10\%)     \\ \hline 
 AlAs    & 360~(360,0\%)                             &  425 (402, -6\% )     \\ \hline
 AlSb    & 323~(323,0\%)                             &  362 (344, -5\% )     \\ \hline
  GaN    & 560~(560\textsuperscript{b},0\%)                             &  878 (750\textsuperscript{b},-17\%)                                    \\ \hline
  GaP    & 366~(366,0 \%)                            &  421 (403, -4\%)    \\ \hline
 GaAs    & 273~(273,0\%)                             &  321 (296, -8\%)     \\ \hline
 GaSb    & 231~(231,0\%)                             &  262 (240,-9\% )     \\ \hline
  InN    & 478~(478,0\%)                             &  724 (694, -4\%)                                    \\ \hline
  InP    & 307~(307,0\%)                             &  369 (350, -6\% )                                  \\ \hline
 InAs    & 217~(217,0\%)                             &  260 (240, -8\%)                                    \\ \hline
 InSb    & 180~(180,0\%)                             &  203 (192, -6\% )     \\ \hline
\end{tabular}
\caption{Value of transverse and longitudinal optical phonon frequency at the $\Gamma$ point, $\omega_{\textrm{\tiny{LO}}}$, predicted from Coulombic VFF potential with effective charges 
determined by elastic and inner elastic properties. Experimental values and 
percentage difference are given in brackets. Apart from AlN, a=Ref.~\onlinecite{KiLa96}, and GaN, b=Ref.~\onlinecite{KaWa98},  all experimental
values are the same as those in Table~\ref{tab:Conv_E11}. } \label{tab:StableCoulwLO} 
\end{table}

With this potential, all elastic properties input are reproduced exactly, as is $\omega_{\textrm{\tiny{TO}}}$,
through the inner elastic constant $E_{11}$. The force constants obtained using eqs.~(\ref{eq:S_E11}) and eqs.~(\ref{Cforce1}-\ref{Cforce4}), 
for selected III-V materials, are shown in Table~\ref{tab:FreeCoulforce_consts}. Of particular note in 
Table~\ref{tab:FreeCoulforce_consts} is the much larger screened Coulomb parameter $S=\frac{Z^{*2}}{\epsilon_{r}}$
compared to the conventional parameterisation shown in Table~\ref{tab:ConvCoulforce_consts}. 
We attribute this to the greater importance of short-ranged Coulomb interactions over long-ranged interactions
for the stabilisation of the crystal with respect to internal strains. Interactions between closer atoms will have fewer atoms 
and electrons between them to screen the field, 
and prioritising these interactions will manifest as a larger $S$ in the potential. In addition, 
it is possible that longer range forces other than the Coulomb interaction are being effectively incorporated into this parameter.
Either way, the potential represents a significant improvement in the description of the elastic properties
of the highly ionic ZB structured materials.

Table~\ref{tab:StableCoulwLO} shows a comparison
of calculated $\omega_{\textrm{\tiny{LO}}}$ versus previous theory and experimental values. Comparing with Table~\ref{tab:Conv_E11},
we find that the free parameterisation offers a universal improvement over the conventional parameterisation. 
Being directly fitted to $E_{11}$ it reproduces $\omega_{\textrm{\tiny{TO}}}$ exactly, 
and for $\omega_{\textrm{\tiny{LO}}}$, to which it was not fit, it also performs considerably better.

In the next section, we will perform a further benchmarking 
of each of the potentials. We first benchmark the models 
against first principles DFT relaxations. We find the agreement 
between the VFF relaxed atomic positions and those obtained from DFT is good, and that
again, the new effective charge parameterisation produces the best results.
We then compare their relative performances in the 
calculation of phonon spectra, where we show best overall agreement with experiment is obtained for 
the third model presented.


\section{Comparison with experimental and \emph{ab-initio} data} \label{sec:Benchmarking}

In this section, we present a benchmarking of the three different potentials. We
first validate the potentials for use as a tool for structural relaxation: we find, using
each potential, the relaxed atomic positions in various InAs/GaAs supercells and compare these positions with those obtained 
from DFT calculations within the local density approximation (LDA). Then, we analyse and compare with experiment the VFF calculated phonon bandstructure of GaAs.
Our choice of GaAs/InAs systems for benchmarking is based on the following considerations: InAs/GaAs heterostructures
are one of the most technolgically relevant semiconductor material systems, widely studied, and grown along various different crystollographic
directions;\cite{Ramzi16} secondly, both InAs and GaAs are ionic materials, with InAs being a material for which the anisotropy factor
is just past the threshold of stability ($A<2$) for the covalent potential, and therefore a system which is a combination of these 
two binary compounds serves
as an ideal test bed for the different variants of the potential.

To benchmark the potentials against first principles structural relaxations, we first parameterise our
VFF using elastic constants from DFT calculations commensurate with those from which the relaxed
atomic positions were determined. While the force constants presented earlier will more accurately 
reproduce the true atomic positions (since the hybrid-functional-DFT elastic constants agree better with
experiment), performing test structure relaxations using HSE DFT is computationally costly, and not necessary. 
When benchmarking, no extra information is gained by making comparisons to a computationally expensive functional.

The elastic constants $C_{ij}$ and the Kleinman parameter, $\zeta$ were calculated
using LDA DFT, using a k-point grid of 16$\times$16$\times$16 and a cutoff
energy of 600 eV and are given in Table~\ref{tab:LDACij}.
These elastic constants were used to parameterise the three
VFF models via eqs.~(\ref{force1}) to (\ref{force4}); eqs.~(\ref{Cforce1}) to (\ref{Cforce4}); 
and (\ref{eq:S}) and (\ref{eq:S_E11}).

\begin{table}[t] 
\centering  
\begin{tabular}{|c|c|c|c|c|c|c|}
\hline
     & $a_{0}$ & $C_{11}$ & $C_{12}$ & $C_{44}$  & $\zeta$ & $E_{11}$ \\ \hline
GaAs & 5.6198  & 115      & 52       & 58        & 0.547   & 34       \\ \hline
InAs & 6.0312  & 85       & 48       & 38        & 0.687   & 23       \\ \hline 
\end{tabular}
\caption{LDA DFT calculated elastic and structural properties of GaAs and InAs. Calculations were
performed on a k-point grid of 16$\times$16$\times$16 and a planewave cutoff energy of 600 eV. $a_{0}$ is in \AA, $C_{ij}$ are in GPa, $\zeta$ is dimensionless, and 
$E_{11}$ is in GPa~\AA$^{-2}$. }  \label{tab:LDACij}
\end{table}

Next, four different supercells have been relaxed using LDA DFT: (i) a simple GaAs/InAs interface along the [001]
crystallographic direction, modelled as a supercell of alternating GaAs/InAs conventional 
unit cells, containing 16 atoms and having unrelaxed dimensions $a_{0},a_{0},2a_{0}$, where $a_{0}=5.6198$~\AA, in the $x$, $y$, and $z$ directions,
respectively; (ii) a (001) quantum well type interface, consisting of a GaAs cubic unit cell, an InAs cubic unit 
cell, and then another GaAs cell, containing 24 atoms and having initial dimensions $a_{0},a_{0},3a_{0}$; (iii) a GaAs/InAs interface along the [111]-direction, consisting
of alternating GaAs/InAs 6-atom unit cells\cite{Miguel_Stress,ScCa11} with the $z$-axis along the [111]- direction, containing
12 atoms and having unrelaxed lattice vectors $a_{1}=(\frac{a_{0}}{\sqrt{2}},0,0)$, $a_{2}=(\frac{a_{0}}{2\sqrt{2}},\frac{\sqrt{3}a_{0}}{2\sqrt{2}},0)$, 
$a_{3}=(0,0,2\sqrt{3}a_{0})$; (iv) a 64 atom GaInAs supercell, consisting of a 2$\times$2$\times$2 
replication of a conventional ZB cell, with In atoms substituted for Ga atoms with a 
probability according to the nominal In content of 25\%. For each of these supercells,
the free energy was minimised until the force on any atom was less than 0.001 eV/\AA. The LDA calculations
were in all cases performed with a cutoff energy of 600 eV, and k-point grid densities of: 12$\times$12$\times$6, 12$\times$12$\times$4,
12$\times$12$\times$5, and 8$\times$8$\times$8, for supercells (i-iv), respectively.

Following the relaxation of each of these supercells using LDA DFT,
the same supercells were relaxed using the three parameterisations of the VFF in the software package GULP.\cite{GULP}
A summary of the comparison between the relaxations produced by these VFF potentials and LDA DFT 
is presented in Table.~\ref{tab:Compare}. 
\begin{table}[t] 
\centering  
\begin{tabular}{|c|c|c|c|c|}
\hline
Supercell              & VFF                 & $\overline{\Delta |\mathbf{a}_{i}|}$(\%)  & $\overline{\Delta r_{ij}}$ (\%) & $\overline{\Delta \theta}$(\%)  \\ \hline
                       & (a)                 & 0.45                & 0.15                & 0.33                \\
  {[001]} GaAs/InAs    & (b)                 & 0.35                & 0.27                & 0.32                \\   
                       & (c)                 & 0.36                & 0.15                & 0.26                \\ \hline \hline
                       & (a)                 & 0.36                & 0.07                & 0.27                \\ 
{[001]} GaAs/InAs/GaAs & (b)                 & 0.28                & 0.19                & 0.27                \\   
                       & (c)                 & 0.26                & 0.11                & 0.22                \\ \hline \hline
                       & (a)                 & -                   & -                    & -                    \\ 
{[111]} GaAs/InAs      & (b)                 & 0.19                & 0.37                 & 0.48                \\   
                       & (c)                 & 0.22                & 0.32                 & 0.48                \\ \hline \hline
                       & (a)                 & 0.03                & 1.94                 & 1.5                  \\ 
 InGaAs alloy          & (b)                 & 0.04                & 0.35                 & 0.37                \\   
                       & (c)                 & 0.05                & 0.25                 & 0.32                \\ \hline \hline
                       & (a)                 & 0.28                & 0.72                 & 0.70                \\ 
 All                   & (b)                 & 0.22                & 0.30                 & 0.36                \\   
                       & (c)                 & 0.22                & 0.21                 & 0.32                \\ \hline \hline
\end{tabular}
\caption{Percentage differences between structural properties of supercells relaxed using LDA DFT, and three
different VFF models. See text for description of supercells; 'All' refers to an averaging of all supercell errors.
VFF (a) is the covalent VFF, (b) is the conventionally Coulombic VFF, and (c) is the freely parameterised Coulombic VFF.
$\overline{\Delta |\mathbf{a}_{i}|}$ denotes the average difference 
in the magnitude of the lattice vectors; $\overline{\Delta r_{ij}}$ is the average difference
in all bondlengths; and $\overline{\Delta \theta}$ is the average difference in angles. }\label{tab:Compare}
\end{table}

Examining first the averaged results presented at the bottom of Table~\ref{tab:Compare}, a trend of increasing 
accuracy in the reproduction of all quantities is seen when progressing from the covalent potential, 
through to the conventionally parameterised ionic potential, to the new free parameterisation of the effective charge. 
This perhaps indicates the importance of accurately 
describing $E_{11}$ for structural relaxations.

Looking in more detail, we find for the covalent potential, that it is able to well relax the two [001] oriented sytems for which
there are no macroscopic shear strains, but it fails completely for the (111) interface and alloy sytems. For the [111]-oriented system, GULP is unable 
to minimise the energy density resulting from the unstable potential. For the alloy supercell GULP is able to achieve a
minimum, owing to the stabilising effect of the GaAs matrix, but the instability of the InAs VFF with respect to
shear strains is manifested in larger errors in bondlengths and angles. 
For the ionic potentials, a good description of the lattice and bond properties is found for 
all systems, and unlike the covalent potentials, there is no increase in the errors for
the [111]-oriented or alloyed structures. 
For all potentials, aside from the unstable covalent potential, the errors in the relaxation of the alloy
supercell are much lower than those in the layered systems. This may be imputed to 
nonlinear strain effects experienced in the sharply interfaced supercells - the 
errors in this case could be reduced by inclusion of anharmonic 
force constants and third order elastic constants.\cite{TaCa19} 


Overall, the agreement between the first-principles and VFF
relaxations of the here-considered supercells is very good, and serves to validate the VFF for
use in larger scale structural relaxations.
\begin{figure}[t]
   \centering
   \includegraphics[width=.4\textwidth]{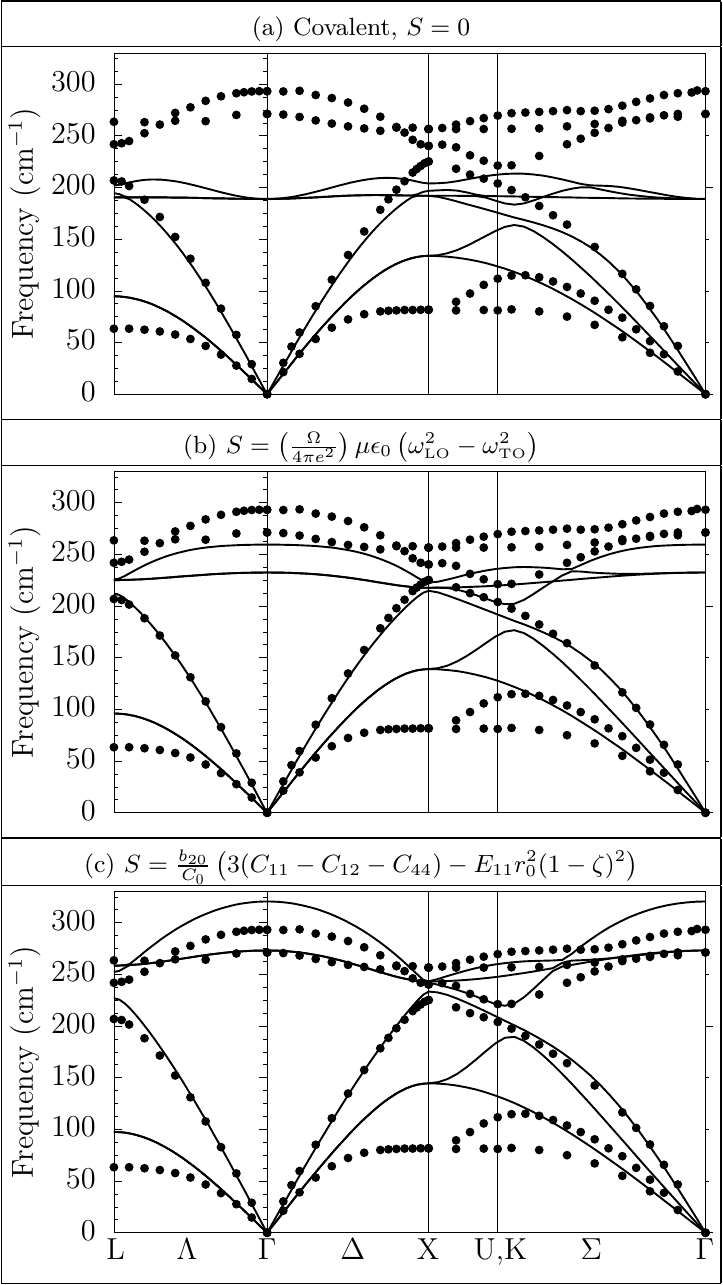}
   \caption{Phonon bandstructure of GaAs calculated using different VFF parameterisations: (a) bandstructure calculated using Covalent VFF 
   with effective charge parameter $S=0$, and force constants described by eqs.~\ref{force1}-\ref{force4}; 
   (b) bandstructure calculated using ionic VFF, with effective charge determined via eq.~\ref{eq:S},
   and force constants determined by eqs.~\ref{Cforce1}-\ref{Cforce4};
   (c) bandstructure calculated using ionic VFF, with effective charges determined 
   via eq.~\ref{eq:S_E11} and force constants given by eqs.~\ref{Cforce1}-\ref{Cforce4}. The filled symbols
   are experimental frequencies taken from Ref.~\onlinecite{Strauch90}.}
   \label{fig:phon}
\end{figure}

Next, the full phonon bandstructure of GaAs, calculated using each parameterisation of
the potential, and determined experimentally,\cite{Strauch90} is shown in Fig.~\ref{fig:phon}.
All three of the parameterisations share a good description of the acoustic modes, near the $\Gamma$-point especially, 
with the description of the longitudinal acoustic modes remaining good
at larger wavevectors. All three potentials share the property that the softening
of the transverse acoustic mode, in the $\Gamma$ to $X$, $L$, and $K$ directions, is not well
described; this is a characteristic feature of nearest neighbour VFFs, and may be remedied,
for example, by inclusion of an angular interaction term which involves four coplanar bonds.\cite{SuIr93,PaLu10,CousinsPhysBI,StSa11,BaWa15}
However, given that our aim is to introduce a potential for simple, accurate, and efficient structural relaxation,
rather than accurate phonon dispersions through the full Brillouin zone, we do not here include this term. 

Looking to the differences between the different potentials, we find that, compared to the other two,
the covalent VFF has larger errors in the longitudinal acoustic modes at large wavevectors, and that
its description of the optical modes is qualitatively and quantitatively significanly inferior to that
of the two ionic potentials; this is to be expected, given the non-negligible ionicity of GaAs.
Comparing the two ionic models, we find that using $E_{11}$ to parameterise the effective charge 
produces a bandstructure which generally agrees better with experiment than that produced
by the potential with a conventionally parameterised effective charge; however, the conventional parameterisation
does produce better agreement with experiment for the longitudinal acoustic branch at $L$. 

Overall, we can conclude
that all potentials reproduce well the acoustic branches near the $\Gamma$-point, while the best agreement with 
experiment throughout the Brillouin zone is obtained by the potential in which 
the effective charge is determined by fitting to the elastic and inner elastic properties.
This shows, in combination with Table~\ref{tab:Compare}, that the new Coulombic parameterisation produces improved relaxation \emph{and}
phonon spectra compared to the conventional parameterisation.

\section{Conclusion} \label{sec:Conclusion}
In conclusion, we have presented a VFF model, based on that originally 
introduced by Musgrave and Pople~\cite{MusPop62} and modified by Martin,\cite{Ma1970} which 
explicitly fits to the often neglected and ill-described 
Kleinman parameter, as well as the three cubic second order elastic constants, of 
which $C_{44}$ is often poorly represented in the popular Keating 
model.\cite{Keating66} Three different parameterisations of the potential were presented: a covalent  
(non-Coulombic) one for non-ionic or weakly ionic materials; and two parameterisations
which include electrostatic forces, in one of which we determine the effective charges via zone-centre
phonon frequencies, while in the second case the effective charges are determined via the static
elastic properties.

The force constants of the model were derived analytically
with explicit expressions given for the force constants in terms of macroscopic 
elastic constants, as well as inner elastic properties which can be measured 
and/or directly calculated using density functional theory. This allows the potential to be used
for a given material without the need for any additional numerical fitting:
once the elastic and related properties of the material are known, the force constants
can be obtained immediately from them by means of the analytic expressions
presented here.

In addition to ease of application, the analytic determination of force
constants also has the advantage that it allows for the a-priori prediction
of properties outside of the determining parameter set of the potential.
This capability allowed for the analysis of the suitability of the potential
for application to different materials. This analysis furnished the result
(general for nearest neighbour VFFs), that a stable non-Coulombic potential which
accurately describes the three cubic elastic constants and Kleinman's internal
strain parameter is not achievable for materials for which the anisotropy factor, $A$, is $< 2$.
The stabilising effect of the Coulomb interaction was first examined based on 
conventional parameterisation in terms of the optical phonon splitting frequency. This parameterisation was found to
stabilise most materials, with the exception of the highly ionic cubic III-N materials, GaN and AlN. 
This instability was remedied by use of a new parameterisation of
the effective charges, which resulted in a potential capable of fully describing 
the elastic energy density of any diamond or zincblende crystal. In 
benchmarking against DFT and experiment, this new paramterisation of the effective charge was shown to produce 
improved phonon spectra and structural relaxations.

The described potential thus offers an efficient, intuitive, and accurate  description
of all classes of zincblende or diamond crystal; with increased accuracy, efficiency
and clarity when compared with machine-learning-based or other complex potentials; and with increased accuracy
at little extra computational cost when compared with the extensively used simpler VFFs predominantly
used for structural relaxation in the literature.

\section*{Acknowledgments}
This work was supported by Science Foundation Ireland (project
numbers 15/IA/3082 and 13/SIRG/2210) and by the European Union 7th
Framework Programme DEEPEN (grant agreement no.: 604416). 

\section{Appendix} \label{appendix1}
We note that eqs.~(14) provide five linear relationships between five macroscopic 
elastic constants ($C_{11}$, $C_{12}$, $C^{0}_{44}$, $D_{14}$ and $E_{11}$) and the
five parameters required in our general VFF model. We can therefore solve these linear
equations directly to obtain expressions for the VFF parameters in terms of macroscopic
elastic properties that can be determined using well established DFT approaches. While 
this is useful, it may be generally preferred to calculate the VFF parameters in terms 
of the experimentally accessible elastic constants, $C_{11}$, $C_{12}$ and $C_{44}$, 
as well as the internal strain parameter $\zeta$, in particular given that an accurate 
description of $\zeta$  is required for an accurate  description of relative atomic 
displacements within a given unit cell. We outline here how the covalent VFF terms 
can be calculated from the linear expressions for $C_{11}$ and $C_{12}$ in eqs.~(\ref{equations}) and from the nonlinear expressions for $C_{44}$
and $\zeta$ in eqs.~(\ref{eq:C44}) and (\ref{eq:Klein}). The method that we describe here can be readily
modified to treat the more general case of the ionic potential with additional 
terms proportional to $SC_{0}$. 

Subtracting $C_{11}$ from $C_{12}$ in eqs.~(\ref{equations}) reveals immediately the unique determination of $k_{\theta}$ 
in terms of $C_{11}$ and $C_{12}$:
\begin{equation} \label{eq:kt}
 k_{\theta} = \frac{2r_{0}}{3\sqrt{3}}\left(C_{11}-C_{12} \right).
\end{equation}
Adding twice $C_{12}$ to $C_{11}$ in eqs.~(\ref{equations}) furnishes a linear 
expression for $k_{rr}$ in terms of $C_{11}$, $C_{12}$, and $k_{r}$:
\begin{equation} \label{eq:krr}
k_{rr} = \frac{2r_0}{3\sqrt{3}}\left(C_{11}+2C_{12} \right)-\frac{k_{r}}{6}.
\end{equation}
Multiplying out eq.~(\ref{eq:Klein}) and utilising eq.~(\ref{eq:krr}), a linear expression relating $k_{r\theta}$ to $k_{r}$ is obtained:
\begin{multline} \label{eq:kr}
 k_{r} = k_{r\theta}\frac{3}{\sqrt{2}}\frac{4\zeta-1}{\zeta-1}-\frac{3k_{\theta}\left(2\zeta+1\right)}{\zeta-1}\\
 +\frac{r_{0}\left(C_{11}+2C_{12} \right)}{\sqrt{2}}
\end{multline}
Having now expressions for $k_{rr}$ in terms of $k_{r}$, and $k_{r\theta}$ in terms of $k_{r}$, the remaining 
equation for $C_{44}$, eq.~(\ref{eq:relax_C44}) can be cast in terms of only $k_{r\theta}$, and known elastic constants. 
Expanding out eq.~(\ref{eq:relax_C44}) we are left with the quadratic equation:
\begin{multline} \label{eq:krt_quad}
 \overbrace{\frac{3}{2r_{0}^{2}}}^{a}k_{r\theta}^{2} + 
 \overbrace{\frac{ 3C_{44}+C^{\prime}\left(1-4\zeta\right)}{\sqrt{6}\left(\zeta-1\right)r_{0}}}^{b} k_{r\theta}
 \\+ \underbrace{\frac{C^{\prime}
 \left(C^{\prime}-3C_{44}+2C^{\prime}\zeta\right)}{9\left(\zeta-1\right)}}_{c} = 0 .
\end{multline} 
This may be solved using the quadratic formula: $k_{r\theta}=\frac{-b\pm\sqrt{b^{2}-4ac}}{2a}$. 
The two solutions then correspond to different values of $k_{r\theta}$, $k_{rr}$, 
and $k_{r}$, with the same $k_{\theta}$.
However, implementation of this formula reveals one of the
solutions to be extraneous, as discussed further below.

Taking the coefficients from eq.~(\ref{eq:krt_quad}), we obtain:
\begin{equation}
 b^{2}-4ac = \frac{3}{2r_{0}^{2}}\frac{\left(C_{44}-C^{\prime} \right)^{2}}{\left(\zeta-1\right)^{2}} \label{eq:b^2-4ac} . 
\end{equation}
Putting this into the quadratic formula and simplifying, we obtain:
\begin{multline} \label{eq:krt_plumin}
 k_{r\theta} = -\frac{4r_{0}}{3\sqrt{6}} \frac{3C_{44}-C^{\prime}\left(1-4\zeta\right) }{\zeta-1} \\
 \pm \frac{4r_{0}^{2}}{3}\sqrt{\frac{3}{2}\left(\frac{C_{44}-C^{\prime} }{\left(\zeta-1\right)r_{0}}\right)^{2}}
\end{multline} 
These two solutions simplify to:
\begin{equation} \label{eq:incorrect}
k^{+}_{r\theta} = \frac{2}{3} \sqrt{\frac{2}{3}} r_{0} C^{\prime}= \sqrt{2}k_{\theta}  ,
\end{equation} 
and, already given in eq.~(\ref{force4}) in Sec.~\ref{sec:Non-Coul} above:
\begin{equation} \label{eq:correct}
 k^{-}_{r\theta} = \frac{r_{0}}{3}\sqrt{\frac{2}{3}}\frac{\left(C_{11}-C_{12}\right)\left(1+2\zeta\right)-3C_{44}}{\zeta-1} .
\end{equation}

By inspection of eq.~(\ref{equations}) we can see
that the extraneous solution is that in eq.~(\ref{eq:incorrect}), which would lead to the 
undefined scenario of $0/0$ in eqs.~(\ref{eq:C44}) and (\ref{eq:Klein}). Furthermore, 
we see that whether this solution is that with the positive or negative root depends 
on whether $C_{44} > C^{\prime}$, equivalent to whether $\frac{A}{2}>1$. In addition, 
we note that these two conditions also govern whether or not the VFF
will be stable against internal strain ($E_{11} >0$); so the result also holds that as the sign of
the extraneous solution changes, so does the sign of $E_{11}$. When this sign change occurs, the
underlying assumption in the derivation of the equations that the energy has been  minimised with
respect to the internal strain becomes invalid.

Thus, the single correct analytic expression for the force constant $k_{r\theta}$ in terms of
the elastic constants and the Kleinman parameter is the right hand solution in eq.~(\ref{eq:correct}). 
Via, eqs.~(\ref{eq:kt}), (\ref{eq:krr}) and (\ref{eq:kr}), we then obtain
the full single set of force constants of eq.~(\ref{force1}-\ref{force4}).

Alternatively, the pitfalls of the extraneous root may be more efficiently circumvented by simply
solving the equation set comprising $C_{11}$, $C_{12}$ and $\zeta$, from eq.~(\ref{equations}) along 
with the rightmost expression of eq.~(\ref{eq:relax_C44}), where the $\zeta$ is not swopped for
its numerical value, but rather left as a known numerical quantity. Choosing this set of equations
a quadratic term in $k_{r\theta}$ never arises, and there is simply a squared $\zeta$, which adds
no extra roots to the equation set.

\bibliographystyle{apsrev4-1}
\bibliography{./Daniel_Tanner_Bibliography}

\end{document}